\documentclass[a4paper,12pt]{article}
\pdfoutput=1

\usepackage{jheppub}
\usepackage[utf8]{inputenc}
\usepackage[english]{babel}

\usepackage{bm}
\usepackage{amsfonts}
\usepackage{mathbbol}
\usepackage{amsthm}
\usepackage{amssymb}
\usepackage{mathrsfs}
\usepackage{latexsym}
\usepackage{commath} 
\usepackage{float}
\usepackage{subfig}
\usepackage{subcaption}
\usepackage{comment}

\usepackage{soul}
\usepackage{comment}
\usepackage{xcolor}

\newcommand{\dd}{\mathrm{d}}
\newcommand{\ii}{\mathrm{i}}
\newcommand{\e}{\mathrm{e}}
\newcommand{\cO}{\mathcal{O}}
\newcommand{\E}{\mathbb{E}}
\DeclareMathOperator{\Vol}{Vol}

\newcommand{\avg}[1]{\left\langle #1 \right\rangle}

\newcommand{\gammaE}{\gamma_{\rm E}}

\title{Excited string states and D-branes from infinite width neural networks}

\author{Dmitry S. Ageev$^{a,b}$ and Yulia A. Ageeva$^{b,c,d}$}

\affiliation{$^{a}$Steklov Mathematical Institute, Russian Academy of Sciences,\\
Gubkin str. 8, 119991 Moscow, Russia\\
  $^{b}$Institute for Theoretical and Mathematical Physics,
  M.V.~Lomonosov Moscow State University, Leninskie Gory 1,
119991 Moscow,
Russia\\
$^{c}$Institute for Nuclear Research of
         the Russian Academy of Sciences,  60th October Anniversary
  Prospect, 7a, 117312 Moscow, Russia\\
 $^{d}$Department of Particle Physics and Cosmology, Physics Faculty, M.V. Lomonosov Moscow State University, Leninskie Gory 1-2, 119991 Moscow, Russia}

\emailAdd{ageev@mi-ras.ru}
\emailAdd{ageeva@inr.ac.ru}

\abstract{
We explore  recent proposal to represent worldsheet string path integrals by integrating over
parameters of a wide random-feature neural network whose output is identified with the embedding field
$X^\mu$. In this paper we extend it focusing on scattering with excited states insertions and for worldsheets with boundaries introducing fixed-feature Gaussian normal-ordering prescription for derivative composites (removing the neural
contact term at finite width), and propose  realization of mixed Neumann/Dirichlet
boundary conditions  interpreted as a neural D$p$-brane.
As concrete outputs, we derive the sphere four-point integrand with a single $(1,1)$ insertion and the disk
four-tachyon amplitude on a D$p$-brane, recovering the expected derivative prefactors, boundary exponents,
and momentum-conservation limits after renormalization.
}

\begin{document}
\begin{flushright}
\end{flushright}
\maketitle


\section{Introduction}

Perturbative string theory is  efficiently formulated in the Polyakov path integral, in which one integrates over worldsheet metrics and embeddings $X^\mu$ to obtain scattering amplitudes \cite{Polyakov:1981,Green:1987sp,Friedan:1985ge}. After gauge fixing, tree-level amplitudes reduce to correlators in a two-dimensional conformal field theory of free bosons (and the $bc$ ghost system), with the kinematics encoded by vertex operators and their operator product expansions; standard accounts may be found in \cite{Polchinski:1998rr,Green:1987sp}. In this language, the foundational open- and closed-string four-point functions are captured by the Koba--Nielsen representation \cite{Koba:1969kh}, reproducing the Veneziano amplitude for open strings \cite{Veneziano:1968yb} and the Virasoro--Shapiro amplitude for closed strings \cite{Virasoro:1969bx,Shapiro:1970} once the moduli are fixed and the ghost contributions are included.

A priori, however, the worldsheet path integral remains a functional integral, and it is natural to ask whether there exist alternative representations that preserve the salient conformal and target-space structures while making the measure and renormalization completely explicit. Recent developments in the theory of random wide neural networks provide precisely such a framework. In particular, classical results on Bayesian neural networks and infinite-width limits show that broad classes of architectures converge to Gaussian processes, yielding a tractable ``free-field'' description in function space \cite{Neal:1996bnn,Williams:1997inf,Lee:2017dgp,Rasmussen:2006gpml,Jacot:2018ntk}. Building on this, the neural-network field theory (NN-FT) program proposes to regard an ensemble of neural networks, together with a probability density on its parameters, as defining a statistical field theory whose correlators are computed directly in parameter space \cite{Halverson:2020nnqft,Demirtas:2023nnft,Halverson:2024cfnn,Ferko:2025qmnn,Frank:2025fsnnft,Ferko:2026axm}. In this perspective, the architecture and its parameter measure play the role of the defining ultraviolet data of the theory, while locality and interactions may be studied through controlled departures from Gaussianity or from strict statistical independence of parameters \cite{Demirtas:2023nnft,Ferko:2026axm}.

A particularly interesting recent step is the observation by Frank and Halverson that bosonic string theory itself admits such a realization: the free-boson worldsheet theory (together with the ghost sector) can be engineered from an infinite-width random-feature ensemble, and the resulting parameter-space computation reproduces the Veneziano and Virasoro--Shapiro amplitudes as neural correlators \cite{Frank:2026nnstring}. Their construction provides a new viewpoint on familiar string-theoretic structures: the logarithmic propagator emerges from an explicit feature-space integral in a scaling regime, the Koba--Nielsen factors arise from a law-of-large-numbers limit of i.i.d.\ features, and momentum conservation is recovered through a controlled infinite-variance limit of an explicit Gaussian zero-mode integral \cite{Frank:2026nnstring,Rahimi:2007rf}.

The aim of the present paper is to extend and explore this neural rewriting of the worldsheet path integral in two directions that are important to string perturbation theory but require additional care in a neural regularization. 

First, we incorporate excited closed-string insertions already at genus zero. In conventional CFT language, the simplest level-$(1,1)$ vertex involves the composite $\partial X^\mu \bar\partial X^\nu e^{ip\cdot X}$, whose definition relies on an implicit normal-ordering prescription that removes coincident-point contractions \cite{Polchinski:1998rr,Green:1987sp}. In a neural ensemble, the same physics reappears as an ultraviolet-sensitive local ``contact'' contribution to $\partial X\,\bar\partial X$ that is not removed by the multiplicative renormalization used for the exponential sector alone. We propose a renormalization scheme adapted to the parameter-space computation: we normal order the derivative composite with respect to the Gaussian output-weight average at fixed random features. This subtraction is exact at finite width and fixed features, and it isolates the physically meaningful part of the operator that contracts against the exponentials, thereby reproducing the expected free-field Ward identities in the large-width limit.

Second, we introduce worldsheets with boundary and mixed Neumann/Dirichlet conditions, i.e.\ D$p$-brane sectors. In the standard worldsheet formulation, D-branes are implemented by boundary conditions on $X^\mu$ together with an overall disk normalization identified with the brane tension \cite{Dai:1989ua,Polchinski:1995mt,Polchinski:1998rr}. We give a neural realization of these boundary conditions using an image-like construction directly at the level of the bulk neural field, yielding boundary fluctuations only along Neumann directions and frozen Dirichlet zero modes at the brane position. As in the conventional treatment, we find that the resulting disk correlators are determined by a boundary logarithmic kernel and are multiplied by a single overall constant, which we interpret as the neural avatar of the D$p$-brane tension.

We calculate  the sphere four-point matter integrand with three tachyons and one renormalized $(1,1)$ insertion, recovering the standard Koba--Nielsen factor multiplied by the familiar rational prefactor produced by contracting $\partial X$ and $\bar\partial X$ against the exponentials, with the correct $\alpha'$-normalization fixed by the neural scaling; and the disk four-tachyon amplitude in a D$p$ sector, recovering the boundary Koba--Nielsen exponents and the emergence of momentum conservation along the brane in the appropriate infinite-variance limit.

The   paper is organized as follows. We first review the neural ensemble for the embedding field and the renormalization logic for exponential vertices following \cite{Frank:2026nnstring,Halverson:2020nnqft,Demirtas:2023nnft}. We then treat excited closed-string insertions by formulating the renormalized composite operator and evaluating the resulting correlator explicitly in parameter space, and we finally turn to the disk and D$p$ boundary conditions, where the image construction and the boundary feature kernels give the open-string amplitude in a parallel way.

\section{Neural ensemble for the embedding field}

The worldsheet field theory we wish to represent is the standard free-boson CFT underlying
tree-level string amplitudes. We work throughout on genus zero, using a complex coordinate
$z$ on the sphere (or its upper-half-plane representative in the disk discussion below).
The target-space embedding field $X^\mu(z,\bar z)$ is modeled as a random-feature neural network
of width $N$ whose parameters are integrated over a fixed ensemble measure. The point of this
representation is not to introduce new dynamics---the field remains Gaussian in the relevant sense---but
to trade the formal worldsheet path integral for an explicit finite-dimensional integral over neural
parameters, with the large-$N$ limit playing the role of the continuum limit.

\vspace{0.5em}
\noindent
We take $X^\mu$ to be a superposition of random cosine features,
\begin{align}
X^\mu(z,\bar z)
&=
\frac{C}{\sqrt{N}}
\sum_{i=1}^N
\frac{a_i^\mu}{|W_i|}
\cos\!\Big(\tfrac{1}{2}(W_i z+\bar W_i\bar z)+c_i\Big),
\label{eq:Xdef}
\\
C^2
&=
\frac{\alpha'(\Lambda^2-\epsilon^2)}{\sigma_a^2}.
\label{eq:Cdef}
\end{align}
where the parameters are i.i.d.\ with
\begin{align}
a_i^\mu &\sim \mathcal N(0,\sigma_a^2),
\qquad
c_i \sim \mathrm{Unif}[-\pi,\pi],
\qquad
W_i \sim \mathrm{Unif}(A_{\Lambda,\epsilon})\subset \mathbb C,
\label{eq:ensemble}
\end{align}
and $A_{\Lambda,\epsilon}$ is the annulus
\begin{equation}
A_{\Lambda,\epsilon}
=
\{W\in \mathbb C:\ \epsilon \le |W|\le \Lambda\},
\qquad
\Vol(A_{\Lambda,\epsilon})=\pi(\Lambda^2-\epsilon^2).
\end{equation}
Also $(\Lambda,\epsilon)$ play the role of ultraviolet and infrared regulators in feature space.
The factor $1/|W|$ is chosen so that the induced two-point kernel reproduces the logarithm in the
intermediate scaling window; this is the neural analogue of the fact that a two-dimensional massless
propagator is logarithmic.\footnote{ Here we also adopt the conventional shorthand
$
\partial \equiv \frac{\partial}{\partial z},
\bar\partial \equiv \frac{\partial}{\partial \bar z}.$
}

\vspace{0.5em}
\noindent
As in the usual worldsheet description, it is useful to separate a translational zero mode.
We therefore introduce an independent Gaussian variable $X_0^\mu$ and define
\begin{equation}
\widetilde X^\mu(z,\bar z) \equiv X^\mu(z,\bar z) + X_0^\mu,
\qquad
X_0^\mu \sim \mathcal N(0,\sigma_0^2).
\label{eq:zeromode}
\end{equation}
The role of this zero mode is standard: it is responsible for momentum conservation in the limit of
infinite effective variance after renormalization. In the present representation, this mechanism is
completely explicit because the $X_0$ integral is elementary and can be done as the first step in any
correlator.

\vspace{0.5em}
\noindent
It will be convenient to factor out the overall scale
\begin{equation}
A
\equiv
\frac{1}{\sigma_a}\sqrt{\frac{\alpha'(\Lambda^2-\epsilon^2)}{N}},
\qquad
A^2\sigma_a^2=\frac{\alpha'(\Lambda^2-\epsilon^2)}{N},
\label{eq:Adef}
\end{equation}
and to write the individual random features as
\begin{equation}
f_i(z)
\equiv
\frac{\cos\!\big(\tfrac{1}{2}(W_i z+\bar W_i\bar z)+c_i\big)}{|W_i|}.
\label{eq:fdef}
\end{equation}
Then the field takes the compact form
\begin{equation}
X^\mu(z,\bar z)=A\sum_{i=1}^N a_i^\mu f_i(z).
\label{eq:Xcompact}
\end{equation}

\vspace{0.5em}
\noindent
The neural-network measure factorizes across neurons and across target-space components:
\begin{equation}
[\dd\mu_X]
=
\prod_{i=1}^N
\left[
\frac{\dd^2W_i}{\Vol(A_{\Lambda,\epsilon})}\,
\frac{\dd c_i}{2\pi}\,
\prod_{\mu=1}^D
\frac{\dd a_i^\mu}{\sqrt{2\pi\sigma_a^2}}\,
\exp\!\Big(-\frac{(a_i^\mu)^2}{2\sigma_a^2}\Big)
\right].
\label{eq:measure}
\end{equation}
 We will  explicitly decorate expectation values when needed, for example
$\avg{\cdots}_a$ for the Gaussian average over the output weights $\{a_i^\mu\}$ at fixed features
$\{(W_i,c_i)\}$, and $\E_{W,c}[\cdots]$ for the remaining expectation over features.

Finally, correlators involving exponential vertex operators require the usual removal of self-contractions.
In the neural language this appears as cutoff-dependent constants originating from coincident-feature
contributions. We handle these by the same physics principle as in the continuum CFT: we introduce
multiplicative renormalization factors $Z(\cdot)$ for bulk exponentials and $Z_o(\cdot)$ for boundary
exponentials, chosen so that all cutoff-dependent constants are removed and the momentum-conservation limit
is well-defined. In what follows, we will keep this renormalization logic explicit while focusing on the
dependence on insertion points, which is universal.

\section{Closed-string four-point function with a single $(1,1)$ insertion on the sphere}


We consider a tree-level closed-string amplitude on the sphere with three tachyonic insertions and one
excited insertion at level $(1,1)$. In the usual free-boson CFT, the relevant integrated vertices are:
\begin{align}
V(p)
&=
g_s\,Z(p)\int \dd^2z\; \exp\!\big(\ii\, p\cdot \widetilde X(z)\big),
\label{eq:bulktachyon}
\end{align}
for tachyons, and
\begin{equation}
V_{\rm exc}(p_1,\zeta)
=
g_s\,Z(p_1)\int \dd^2z_1\;
\zeta_{\mu\nu}\,
\partial X^\mu(z_1)\,\bar\partial X^\nu(z_1)\,
\exp\!\big(\ii\,p_1\cdot \widetilde X(z_1)\big),
\label{eq:exc-raw}
\end{equation}
for the $(1,1)$ insertion with polarization tensor $\zeta_{\mu\nu}$. Here $Z(p)$ denotes the
multiplicative renormalization that implements normal ordering of the exponential sector in the chosen
regularization. We keep the integrated-vertex convention and divide by the residual conformal volume.

In the neural regularization there is one subtlety relative to the textbook expression
\eqref{eq:exc-raw}. The composite $\partial X^\mu \bar\partial X^\nu$ contains a local, ultraviolet-sensitive
contact term that is the neural avatar of the coincident contraction at a single point. If we want a
renormalized operator whose correlators reproduce the standard free-field Ward identities, we must remove
this local term. A clean way to do so, adapted to the structure of the neural ensemble, is to normal-order
the composite with respect to the Gaussian output-weight measure at fixed features. Concretely we define
\begin{equation}
\big[\partial X^\mu \bar\partial X^\nu\big]_{\rm ren}(z)
\equiv
\partial X^\mu(z)\bar\partial X^\nu(z)
-
\avg{\partial X^\mu(z)\bar\partial X^\nu(z)}_{a},
\label{eq:ren-composite}
\end{equation}
where $\avg{\cdots}_a$ denotes the average over the Gaussian weights $\{a_i^\mu\}$ with $(W_i,c_i)$ held fixed.
The renormalized excited vertex is then
\begin{equation}
V^{\rm ren}_{\rm exc}(p_1,\zeta)
=
g_s\,Z(p_1)\int \dd^2z_1\;
\zeta_{\mu\nu}\,
\big[\partial X^\mu \bar\partial X^\nu\big]_{\rm ren}(z_1)\,
\exp\!\big(\ii\,p_1\cdot \widetilde X(z_1)\big).
\label{eq:exc-ren-vertex}
\end{equation}
This subtraction is exact at finite width and fixed features. It removes precisely the unwanted
self-contraction term inside $\partial X\,\bar\partial X$ while leaving intact the physically meaningful
contractions of the derivatives against the exponentials in the correlator.

With these definitions, the four-point amplitude is
\begin{equation}
A^{(4)}_{\rm exc}(p_1,\zeta;p_2,p_3,p_4)
=
\frac{1}{g_s^2}\,\frac{1}{\Vol(SL(2,\mathbb C))}\,
\avg{
V^{\rm ren}_{\rm exc}(p_1,\zeta)\,
\prod_{r=2}^4 V(p_r)
}.
\label{eq:amp-def}
\end{equation}
Expanding the integrated vertices reduces the amplitude to an integral over insertion points of a single
matter correlator. Writing $z_\alpha$ for the four insertion points, we define the reduced integrand
\begin{equation}
\mathcal I_{\rm ren}(z_1,z_2,z_3,z_4)
=
\avg{
\zeta_{\mu\nu}\,
\big[\partial X^\mu\bar\partial X^\nu\big]_{\rm ren}(z_1)\,
\exp\!\Big(\ii\sum_{j=1}^4 p_j\cdot \widetilde X(z_j)\Big)
},
\label{eq:Iren-def}
\end{equation}
so that
\begin{equation}
A^{(4)}_{\rm exc}
=
\frac{g_s^2}{\Vol(SL(2,\mathbb C))}
\int \prod_{\alpha=1}^4 \dd^2 z_\alpha\;
\Big(\prod_{r=1}^4 Z(p_r)\Big)\;
\mathcal I_{\rm ren}(z_1,z_2,z_3,z_4).
\label{eq:amp-int}
\end{equation}


In the neural representation, the correlator \eqref{eq:Iren-def} is an explicit integral over neural
parameters and the zero mode:
\begin{align}
&\mathcal I_{\rm ren}
=\int[\dd\mu_X]
\int \dd^D X_0\;
\frac{1}{(2\pi\sigma_0^2)^{D/2}}
\exp\!\Big(-\frac{X_0^2}{2\sigma_0^2}\Big)\,
\zeta_{\mu\nu}\,
\big[\partial X^\mu\bar\partial X^\nu\big]_{\rm ren}(z_1)\,\nonumber\\
&\times\exp\!\Big(\ii\sum_{j=1}^4 p_j\cdot (X(z_j)+X_0)\Big).
\label{eq:Iren-param}
\end{align}
The advantage of this form is that it makes the order of operations completely concrete: we can first
do the $X_0$ integral, then the Gaussian integrals over the output weights $a_i^\mu$ (with the subtraction
\eqref{eq:ren-composite} built in), and only at the end perform the feature average over $(W_i,c_i)$ and
the large-width limit.


We now carry out the computation in the natural sequence.
The insertion $\big[\partial X^\mu\bar\partial X^\nu\big]_{\rm ren}(z_1)$ does not depend on $X_0$, so the
zero mode appears only through $(\sum_j p_j)\cdot X_0$. Completing the square gives
\begin{align}
\int \dd^D X_0\;
\frac{1}{(2\pi\sigma_0^2)^{D/2}}
\exp\!\Big(-\frac{X_0^2}{2\sigma_0^2}+\ii(\sum_j p_j)\cdot X_0\Big)
&=
\exp\!\Big(-\frac{\sigma_0^2}{2}\Big(\sum_{j=1}^4 p_j\Big)^2\Big).
\label{eq:X0bulk}
\end{align}
Therefore
\begin{align}
\mathcal I_{\rm ren}
&=
\exp\!\Big(-\frac{\sigma_0^2}{2}\Big(\sum_{j=1}^4 p_j\Big)^2\Big)\,
\int[\dd\mu_X]\;
\zeta_{\mu\nu}\,
\big[\partial X^\mu\bar\partial X^\nu\big]_{\rm ren}(z_1)\,
\exp\!\Big(\ii\sum_{j=1}^4 p_j\cdot X(z_j)\Big).
\label{eq:Iren-afterX0}
\end{align}
At fixed features $(W_i,c_i)$ the dependence on $a_i^\mu$ is entirely Gaussian. Introduce the feature-smeared
momenta
\begin{equation}
S_i^\mu
\equiv
\sum_{j=1}^4 p_j^\mu\, f_i(z_j),
\label{eq:Sdef}
\end{equation}
so that, using \eqref{eq:Xcompact},
\begin{equation}
\exp\!\Big(\ii\sum_{j=1}^4 p_j\cdot X(z_j)\Big)
=
\exp\!\Big(\ii A\sum_{i=1}^N a_i\cdot S_i\Big).
\label{eq:exp-linear}
\end{equation}
Explicitly differentianting the field we get
\begin{equation}
\partial X^\mu(z_1)\bar\partial X^\nu(z_1)
=
A^2\sum_{i,k=1}^N a_i^\mu a_k^\nu\,
\partial f_i(z_1)\bar\partial f_k(z_1).
\label{eq:deriv-bilinear}
\end{equation}
such that fixed-feature $a$-vacuum contraction is immediate:
\begin{equation}
\avg{a_i^\mu a_k^\nu}_a=\sigma_a^2\,\delta^{\mu\nu}\delta_{ik},
\end{equation}
and the subtraction in \eqref{eq:ren-composite} gives
\begin{equation}
\big[\partial X^\mu\bar\partial X^\nu\big]_{\rm ren}(z_1)
=
A^2\sum_{i,k=1}^N
\big(a_i^\mu a_k^\nu-\sigma_a^2\delta^{\mu\nu}\delta_{ik}\big)\,
\partial f_i(z_1)\bar\partial f_k(z_1).
\label{eq:ren-in-features}
\end{equation}

\vspace{0.5em}
\noindent
Insert \eqref{eq:ren-in-features} and \eqref{eq:exp-linear} into \eqref{eq:Iren-afterX0} and perform the
Gaussian average over the weights at fixed features. For a
single $D$-component Gaussian vector $a^\mu\sim \mathcal N(0,\sigma_a^2)$ we have
\begin{equation}
\avg{\exp(\ii\,t\cdot a)}=\exp\!\Big(-\frac{\sigma_a^2}{2}\,t^2\Big),
\qquad
t^2\equiv t\cdot t,
\label{eq:genfunc}
\end{equation}
with 
\eqref{eq:genfunc} twice,
\begin{equation}
\avg{a^\mu a^\nu\,\exp(\ii\,t\cdot a)}
=
\big(\sigma_a^2\delta^{\mu\nu}-\sigma_a^4\,t^\mu t^\nu\big)\,
\exp\!\Big(-\frac{\sigma_a^2}{2}\,t^2\Big).
\label{eq:a2gen}
\end{equation}
Because the neural measure factorizes over neurons, the multi-neuron expectation is obtained by applying
\eqref{eq:genfunc}--\eqref{eq:a2gen} neuron by neuron with $t_i=A S_i$. One finds the compact result
\begin{align}
\avg{a_i^\mu a_k^\nu\,\exp\!\Big(\ii A\sum_{\ell=1}^N a_\ell\cdot S_\ell\Big)}_a
&=
\big(\sigma_a^2\delta^{\mu\nu}\delta_{ik}-A^2\sigma_a^4 S_i^\mu S_k^\nu\big)\,
\exp\!\Big(-\frac{A^2\sigma_a^2}{2}\sum_{\ell=1}^N S_\ell^2\Big),
\label{eq:multi-a2}
\\[0.3em]
\avg{\exp\!\Big(\ii A\sum_{\ell=1}^N a_\ell\cdot S_\ell\Big)}_a
&=
\exp\!\Big(-\frac{A^2\sigma_a^2}{2}\sum_{\ell=1}^N S_\ell^2\Big).
\label{eq:multi-a0}
\end{align}
The whole point of the subtraction \eqref{eq:ren-in-features} is that it cancels the $\sigma_a^2\delta_{ik}$
term in \eqref{eq:multi-a2} exactly, leaving
\begin{equation}
\avg{
\big(a_i^\mu a_k^\nu-\sigma_a^2\delta^{\mu\nu}\delta_{ik}\big)\,
\exp\!\Big(\ii A\sum_{\ell} a_\ell\cdot S_\ell\Big)
}_a
=
\big(-A^2\sigma_a^4 S_i^\mu S_k^\nu\big)\,
\exp\!\Big(-\frac{A^2\sigma_a^2}{2}\sum_{\ell=1}^N S_\ell^2\Big).
\label{eq:subtracted-avg}
\end{equation}
Inserting this into \eqref{eq:Iren-afterX0} we obtain an exact finite-width expression after the Gaussian
weight integration:
\begin{align}
\mathcal I_{\rm ren}
&=
\exp\!\Big(-\frac{\sigma_0^2}{2}\Big(\sum_{j=1}^4 p_j\Big)^2\Big)\,\nonumber\\
&\times\E_{W,c}\Bigg[
-A^4\sigma_a^4\,\zeta_{\mu\nu}
\Big(\sum_{i=1}^N \partial f_i(z_1)\,S_i^\mu\Big)
\Big(\sum_{k=1}^N \bar\partial f_k(z_1)\,S_k^\nu\Big)\,
\exp\!\Big(-\frac{A^2\sigma_a^2}{2}\sum_{\ell=1}^N S_\ell^2\Big)
\Bigg].
\label{eq:Iren-aftera}
\end{align}
At this stage the only remaining randomness is in the i.i.d.\ features $(W_i,c_i)$.

\vspace{0.5em}
\noindent
The exponent in \eqref{eq:Iren-aftera} has the same structure as in the pure tachyon sector. Define
\begin{equation}
Q_i
\equiv
\sum_{r,s=1}^4 (p_r\cdot p_s)\,f_i(z_r)f_i(z_s),
\label{eq:Qdef}
\end{equation}
so that
\begin{equation}
\sum_{\ell=1}^N S_\ell^2
=
\sum_{\ell=1}^N Q_\ell.
\end{equation}
Using \eqref{eq:Adef}, we can rewrite the exponential as
\begin{equation}
\exp\!\Big(-\frac{A^2\sigma_a^2}{2}\sum_{\ell=1}^N S_\ell^2\Big)
=
\exp\!\Big(-\frac{\alpha'(\Lambda^2-\epsilon^2)}{2N}\sum_{\ell=1}^N Q_\ell\Big).
\label{eq:expQ}
\end{equation}
To streamline notation, define the neuron-level quantities
\begin{equation}
g_i^\mu \equiv \partial f_i(z_1)\,S_i^\mu,
\qquad
\bar g_i^\nu \equiv \bar\partial f_i(z_1)\,S_i^\nu,
\qquad
U^\mu\equiv \sum_{i=1}^N g_i^\mu,
\qquad
\bar U^\nu\equiv \sum_{i=1}^N \bar g_i^\nu,
\label{eq:gUdef}
\end{equation}
and the $N$-dependent tilt parameter
\begin{equation}
\alpha_N \equiv \frac{\alpha'(\Lambda^2-\epsilon^2)}{2N}.
\end{equation}
Then the feature average in \eqref{eq:Iren-aftera} is of the form
\begin{equation}
\E_{W,c}\Big[U^\mu \bar U^\nu\,\exp\!\Big(-\alpha_N\sum_{\ell=1}^N Q_\ell\Big)\Big].
\label{eq:UUform}
\end{equation}
Because the features are i.i.d.\ across neurons, the double sum in $U^\mu\bar U^\nu$ can be decomposed into
terms with $i\neq k$ and $i=k$. Introducing single-neuron expectations (where the expectation is over one
copy of $(W,c)$),
\begin{align}
\Phi_N &\equiv \E\!\big[\e^{-\alpha_N Q}\big],
&
G_N^\mu &\equiv \E\!\big[g^\mu \e^{-\alpha_N Q}\big],
&
\bar G_N^\nu &\equiv \E\!\big[\bar g^\nu \e^{-\alpha_N Q}\big],
&
H_N^{\mu\nu} &\equiv \E\!\big[g^\mu\bar g^\nu \e^{-\alpha_N Q}\big],
\label{eq:single-neuron-moments}
\end{align}
one finds the exact identity
\begin{equation}
\E\Big[U^\mu\bar U^\nu\,\e^{-\alpha_N\sum_{\ell=1}^N Q_\ell}\Big]
=
N(N-1)\,G_N^\mu \bar G_N^\nu\,\Phi_N^{N-2}
+
N\,H_N^{\mu\nu}\,\Phi_N^{N-1}.
\label{eq:combinatorics}
\end{equation}
This is simply the bookkeeping statement that for $i\neq k$ the two insertions live on two independent
neurons (hence the product $G_N\bar G_N$ and $\Phi_N^{N-2}$ from the remaining $N-2$ neurons), while for
$i=k$ they sit on the same neuron (hence $H_N$ and $\Phi_N^{N-1}$).

Now recall from \eqref{eq:Iren-aftera} that the prefactor is $A^4\sigma_a^4\sim N^{-2}$. Therefore:
the first term in \eqref{eq:combinatorics} scales as $N^2\times N^{-2}\sim \cO(1)$ and survives at large
width, whereas the second term scales as $N\times N^{-2}\sim \cO(1/N)$ and is suppressed. Moreover, since
$\alpha_N=\cO(1/N)$, the tilt in $G_N$ is subleading:
\begin{equation}
G_N^\mu = \E[g^\mu] + \cO(1/N),
\qquad
\bar G_N^\nu = \E[\bar g^\nu] + \cO(1/N),
\label{eq:Glimit}
\end{equation}
while $\Phi_N^N$ must be kept nonperturbatively, exactly as in the tachyon computation.

Putting these observations together, the $N\to\infty$ limit of \eqref{eq:Iren-aftera} reduces to (at $N\to\infty$)
\begin{align}
\mathcal I_{\rm ren}
&=
\exp\!\Big(-\frac{\sigma_0^2}{2}\Big(\sum_{j=1}^4 p_j\Big)^2\Big)\,
\Big(-\alpha'^2(\Lambda^2-\epsilon^2)^2\Big)\,
\zeta_{\mu\nu}\,\E[g^\mu]\,\E[\bar g^\nu]\,
\lim_{N\to\infty}\Phi_N^N.
\label{eq:Iren-largeN-structure}
\end{align}
What remains is to evaluate the universal kernel encoded in $\Phi_N^N$ and the derivative kernels encoded in
$\E[g]$ and $\E[\bar g]$.

The large-$N$ limit of $\Phi_N^N$ is controlled by $\E[Q]$:
\begin{equation}
\Phi_N
=
1-\alpha_N\,\E[Q]+\cO(N^{-2}),
\end{equation}
such that
\begin{equation}
\label{eq:PhiNlimit}
\Phi_N^N \xrightarrow[N\to\infty]{}\exp\!\Big(-\frac{\alpha'}{2}(\Lambda^2-\epsilon^2)\,\E[Q]\Big).
\end{equation}
The expectation $\E[Q]$ is built from the two-feature correlator $\E[f(z)f(w)]$. At fixed $W$,
averaging over the random phase $c$ gives
\begin{equation}
\E_c\big[\cos(u+c)\cos(v+c)\big]=\tfrac12\cos(u-v).
\end{equation}
For $z\neq w$, this yields 
\begin{equation}
\E_{c}\big[f(z)f(w)\big]
=
\frac{1}{2}\,\E_c\!\left[\frac{\cos\!\big(\tfrac12(W(z-w)+\bar W(\bar z-\bar w))\big)}{|W|^2}\right]
=
\frac{1}{2}\,\frac{\cos\!\big(\mathrm{Re}(W(z-w))\big)}{|W|^2},
\end{equation}
and averaging $W$ uniformly over the annulus gives a radial Bessel integral. Writing $W=\rho\,\e^{\ii\theta}$,
rotational invariance yields
\begin{equation}
\E_{W,c}\big[f(z)f(w)\big]
=
\frac{1}{\Lambda^2-\epsilon^2}\int_{\epsilon}^{\Lambda}\frac{\dd\rho}{\rho}\,J_0(\rho|z-w|),
\label{eq:bulk-ff-bessel}
\end{equation}
up to an additive constant independent of $z-w$ (coming from the coincident part that is removed by
$Z(p)$ in physical amplitudes). In the intermediate scaling window
\begin{equation}
\frac{1}{\Lambda}\ll |z-w|\ll \frac{1}{\epsilon},
\label{eq:scaling-window}
\end{equation}
the integral \eqref{eq:bulk-ff-bessel} behaves as
\begin{equation}
\int_{\epsilon}^{\Lambda}\frac{\dd\rho}{\rho}\,J_0(\rho|z-w|)
=
-\log|z-w| + \text{(constant)} + \cO(\epsilon|z-w|,\,(\Lambda|z-w|)^{-1}).
\label{eq:bulk-bessel-asym}
\end{equation}
Accordingly,
\begin{equation}
\E_{W,c}\big[f(z)f(w)\big]
=
-\frac{1}{\Lambda^2-\epsilon^2}\log|z-w| + \text{(constant)} + \cdots,
\qquad (z\neq w).
\label{eq:bulk-ff-log}
\end{equation}
Using \eqref{eq:Qdef}, the separation-dependent part of $\E[Q]$ is therefore
\begin{equation}
\E[Q]
=
-\frac{1}{\Lambda^2-\epsilon^2}\sum_{r\neq s}(p_r\cdot p_s)\log|z_r-z_s| + \text{(constant)}.
\label{eq:EQbulk}
\end{equation}
Inserting \eqref{eq:EQbulk} into \eqref{eq:PhiNlimit} produces the standard Koba--Nielsen factor
\begin{equation}
\lim_{N\to\infty}\Phi_N^N
=
\prod_{r<s}|z_r-z_s|^{\alpha' p_r\cdot p_s}\times \text{(constant)}.
\label{eq:KNbulk}
\end{equation}
The constant is cutoff-dependent and is removed, as usual, by the product of renormalization factors
$\prod_r Z(p_r)$; it can be equivalently absorbed into a renormalization of the zero-mode variance,
$\sigma_0^2\to\sigma_{\rm eff}^2$, whose precise form is fixed by the same tachyon-sector normalization.

\vspace{0.5em}
\noindent
We now evaluate the remaining factors $\E[g^\mu]$ and $\E[\bar g^\nu]$. By \eqref{eq:gUdef} and
\eqref{eq:Sdef},
\begin{equation}
\E[g^\mu]
=
\sum_{r=1}^4 p_r^\mu\,\E\big[\partial f(z_1)f(z_r)\big],
\qquad
\E[\bar g^\nu]
=
\sum_{s=1}^4 p_s^\nu\,\E\big[\bar\partial f(z_1)f(z_s)\big].
\label{eq:Egbulk}
\end{equation}
For $r\neq 1$, the correlator $\E[f(z_1)f(z_r)]$ depends on $z_1$ only through $z_1-z_r$ and we may
differentiate \eqref{eq:bulk-ff-log}. Using $\partial_{z}\log|z|=1/(2z)$, we obtain
\begin{equation}
\E\big[\partial f(z_1)f(z_r)\big]
=
-\frac{1}{\Lambda^2-\epsilon^2}\,\partial_{z_1}\log|z_1-z_r|
=
-\frac{1}{2(\Lambda^2-\epsilon^2)}\,\frac{1}{z_1-z_r},
\qquad r\neq 1,
\label{eq:dfkernel}
\end{equation}
and similarly
\begin{equation}
\E\big[\bar\partial f(z_1)f(z_s)\big]
=
-\frac{1}{2(\Lambda^2-\epsilon^2)}\,\frac{1}{\bar z_1-\bar z_s},
\qquad s\neq 1.
\label{eq:dbarkernel}
\end{equation}
The $r=1$ term vanishes already at the $c$-average level (it is proportional to $\E_c[\sin u\,\cos u]=0$),
so at leading order we may write the sums over $r,s$ as sums over $\{2,3,4\}$:
\begin{equation}
\E[g^\mu]
=
-\frac{1}{2(\Lambda^2-\epsilon^2)}\sum_{r=2}^4 \frac{p_r^\mu}{z_1-z_r},
\qquad
\E[\bar g^\nu]
=
-\frac{1}{2(\Lambda^2-\epsilon^2)}\sum_{s=2}^4 \frac{p_s^\nu}{\bar z_1-\bar z_s}.
\label{eq:Egfinal}
\end{equation}

\vspace{0.5em}
\noindent
Combining \eqref{eq:Iren-largeN-structure}, \eqref{eq:KNbulk}, and \eqref{eq:Egfinal}, and using that
$-\alpha'^2(\Lambda^2-\epsilon^2)^2$ cancels the $(\Lambda^2-\epsilon^2)^{-2}$ from the derivative kernels,
we arrive at the large-width expression for the renormalized matter correlator:
\begin{align}
\mathcal I_{\rm ren}(z_1,z_2,z_3,z_4)
&=
\exp\!\Big(-\frac{\sigma_0^2}{2}\Big(\sum_{j=1}^4 p_j\Big)^2\Big)\,
\left[
-\frac{\alpha'^2}{4}\,
\zeta_{\mu\nu}\,
\Big(\sum_{r=2}^4\frac{p_r^\mu}{z_1-z_r}\Big)
\Big(\sum_{s=2}^4\frac{p_s^\nu}{\bar z_1-\bar z_s}\Big)
\right]\nonumber\\
&\times\prod_{r<s}|z_r-z_s|^{\alpha' p_r\cdot p_s},
\label{eq:Iren-final-bulk}
\end{align}
with the understood choice of $Z(p_r)$ that removes the cutoff-dependent self-contraction constants and
produces momentum conservation in the $\sigma_{\rm eff}\to\infty$ limit.

With \eqref{eq:Iren-final-bulk} the amplitude is given by \eqref{eq:amp-int}.
At the level of the integrand, the effect of the $(1,1)$ insertion is exactly the familiar rational
prefactor dictated by the free-boson OPE. The role of the neural subtraction \eqref{eq:ren-composite} is
conceptually sharp: it removes the local $\partial X\,\bar\partial X$ self-contraction at finite width and
fixed features, so that the remaining contribution comes purely from contractions of the derivatives with
the exponentials, as required by Ward identities.

\vspace{0.5em}
\noindent
{\bf A quick gauge-invariance check.}
A standard consistency test is the decoupling of pure-gauge polarizations. Contracting the holomorphic
index with $p_1^\mu$ and using momentum conservation inside the Koba--Nielsen factor gives
\begin{equation}
p_{1\mu}\sum_{r=2}^4\frac{p_r^\mu}{z_1-z_r}
=
\frac{2}{\alpha'}\,\partial_{z_1}\log\!\Big(\prod_{r=2}^4 |z_1-z_r|^{\alpha' p_1\cdot p_r}\Big),
\end{equation}
so for $\zeta_{\mu\nu}=p_{1\mu}\xi_\nu$ the integrand becomes a total derivative in $z_1$ on the sphere,
reproducing the expected decoupling in the integrated amplitude.

\section{Neural D$p$-branes and open-string four-point amplitudes on the disk}

We now turn to an example where the worldsheet itself has a boundary: the disk amplitude for open-string
tachyon scattering in the presence of D$p$ boundary conditions. The logic parallels the previous section:
we first pose the disk amplitude in standard string language, then present the neural representation of
mixed Neumann/Dirichlet conditions via an image-like construction, and finally evaluate the correlator in
the large-width limit to recover the boundary Koba--Nielsen factor and the expected momentum-conservation
structure along the brane.

\subsection{String-theory problem on the disk}

We represent the disk by the upper half-plane
\begin{equation}
\Sigma=\{z=x+\ii y:\ y\ge 0\},
\qquad
\partial\Sigma=\mathbb R,
\end{equation}
with residual conformal symmetry $SL(2,\mathbb R)$. A D$p$-brane is implemented by Neumann boundary
conditions along $p+1$ worldvolume directions and Dirichlet boundary conditions in the transverse space.
We split target indices as
\begin{equation}
\mu=(a,i),
\qquad
a=1,\dots,d_\parallel,\quad d_\parallel=p+1,
\qquad
i=1,\dots,d_\perp,\quad d_\perp=D-p-1,
\end{equation}
and denote the brane position in the transverse space by $Y^i$. For an open-string momentum
$p=(k,q)$ with $k\in\mathbb R^{d_\parallel}$ and $q\in\mathbb R^{d_\perp}$, the integrated boundary tachyon
vertex is
\begin{equation}
V_o(p)
=
\sqrt{g_s}\,Z_o(k)\int_{\mathbb R}\dd x\;\exp\!\big(\ii\,p\cdot \widetilde X_{Dp}(x)\big),
\label{eq:open-vertex}
\end{equation}
where $\widetilde X_{Dp}$ is the D$p$-brane embedding restricted to the boundary. The overall $\sqrt{g_s}$
is the standard open-string normalization in our convention.

The four-point disk amplitude is written, in the same gauge-fixed integrated-vertex convention as before,
as
\begin{equation}
A^{(4)}_{Dp}(p_1,p_2,p_3,p_4;Y)
=
\frac{1}{g_s}\,\frac{1}{\Vol(SL(2,\mathbb R))}\,
\Big\langle\!\Big\langle
\prod_{r=1}^4 V_o(p_r)
\Big\rangle\!\Big\rangle_{Dp}.
\label{eq:open-amp-def}
\end{equation}
Here $\langle\!\langle\cdots\rangle\!\rangle_{Dp}$ denotes the physical disk correlator in the D$p$ sector.
In the neural ensemble it is useful to separate a universal overall normalization:
\begin{equation}
\Big\langle\!\Big\langle \mathcal O \Big\rangle\!\Big\rangle_{Dp}
\equiv
T_p^{\rm NN}\,\Big\langle \mathcal O \Big\rangle_{Dp},
\qquad
\Big\langle 1\Big\rangle_{Dp}=1,
\label{eq:TpNN}
\end{equation}
where $T_p^{\rm NN}$ is a constant common to all disk amplitudes with D$p$ boundary conditions. It plays
the role of a neural D$p$-brane tension. In practice it can be fixed by matching a single chosen disk
one-point function (for example a closed-string probe such as the graviton/dilaton) to a normalization
convention, and then used universally for all other disk correlators.

\subsection{Neural image construction for mixed boundary conditions}

We begin from the bulk neural field $X^\mu(z,\bar z)$ on the plane as defined in
\eqref{eq:Xdef}--\eqref{eq:measure}. On the upper half-plane it is natural to enforce Neumann and Dirichlet
conditions by forming image combinations:
\begin{align}
X_N^\mu(z,\bar z)
&\equiv
\frac{1}{\sqrt 2}\Big(X^\mu(z,\bar z)+X^\mu(\bar z,z)\Big),
&
X_D^\mu(z,\bar z)
&\equiv
\frac{1}{\sqrt 2}\Big(X^\mu(z,\bar z)-X^\mu(\bar z,z)\Big).
\label{eq:imagefields}
\end{align}
On the boundary $z=\bar z=x$ one has $X_D^\mu(x,x)=0$ identically, while $X_N^\mu(x,x)=\sqrt2\,X^\mu(x)$.
The mixed D$p$ embedding is then implemented by
\begin{equation}
X_{Dp}^a(z,\bar z)\equiv X_N^a(z,\bar z),
\qquad
X_{Dp}^i(z,\bar z)\equiv Y^i+X_D^i(z,\bar z),
\label{eq:Xp}
\end{equation}
so that on the boundary
\begin{equation}
X_{Dp}^a(x)=X_N^a(x),
\qquad
X_{Dp}^i(x)=Y^i.
\end{equation}
Thus only the brane-parallel components fluctuate at $\partial\Sigma$, as required.

Restricting the feature \eqref{eq:fdef} to the boundary $z=x$ gives
\begin{equation}
f_i(x)\equiv f_i(z=x)=\frac{\cos\!\big(\mathrm{Re}(W_i)\,x+c_i\big)}{|W_i|},
\label{eq:boundary-feature}
\end{equation}
and therefore
\begin{equation}
X^\mu(x)=A\sum_{i=1}^N a_i^\mu f_i(x),
\qquad
X_N^\mu(x)=\sqrt2\,X^\mu(x)\equiv A_o\sum_{i=1}^N a_i^\mu f_i(x),
\qquad
A_o\equiv \sqrt2\,A.
\label{eq:boundary-field}
\end{equation}

The boundary zero mode is integrated only along Neumann directions. We therefore define
\begin{equation}
\widetilde X_{Dp}^a(x)=X_{Dp}^a(x)+X_0^a,
\qquad
X_0^a\sim \mathcal N(0,\sigma_0^2),
\qquad
\widetilde X_{Dp}^i(x)=Y^i,
\label{eq:open-zeromode}
\end{equation}
so that for $p=(k,q)$,
\begin{equation}
p\cdot \widetilde X_{Dp}(x)=k\cdot (X_N(x)+X_0)+q\cdot Y.
\label{eq:pXopen}
\end{equation}
In particular, each boundary vertex \eqref{eq:open-vertex} carries a phase $\e^{\ii q\cdot Y}$ from the
Dirichlet directions.


Expanding the boundary integrals in \eqref{eq:open-amp-def} yields
\begin{align}
A^{(4)}_{Dp}
&=
\frac{T_p^{\rm NN}}{g_s}\,\frac{1}{\Vol(SL(2,\mathbb R))}\,
\Big(\prod_{r=1}^4 Z_o(k_r)\Big)\,g_s^2\,
\e^{\ii Y\cdot\sum_{r=1}^4 q_r}
\int \prod_{r=1}^4 \dd x_r\;
\mathcal I_o(x_1,x_2,x_3,x_4),
\label{eq:open-expanded}
\end{align}
where the reduced matter correlator is
\begin{equation}
\mathcal I_o(x_1,x_2,x_3,x_4)
=
\Big\langle
\exp\!\Big(\ii\sum_{r=1}^4 k_r\cdot (X_N(x_r)+X_0)\Big)
\Big\rangle_{Dp}.
\label{eq:Io-def}
\end{equation}
As before, we now write this as an explicit parameter-space integral and evaluate it step by step.

\vspace{0.5em}
\noindent
Using \eqref{eq:open-zeromode}, the $X_0$ integral is Gaussian in $d_\parallel=p+1$ dimensions:
\begin{align}
\mathcal I_o
&=
\int[\dd\mu_X]\int \dd^{d_\parallel}X_0\;
\frac{1}{(2\pi\sigma_0^2)^{d_\parallel/2}}
\exp\!\Big(-\frac{X_0^2}{2\sigma_0^2}\Big)\,
\exp\!\Big(\ii\sum_{r=1}^4 k_r\cdot (X_N(x_r)+X_0)\Big)
\nonumber\\
&=
\exp\!\Big(-\frac{\sigma_0^2}{2}\Big(\sum_{r=1}^4 k_r\Big)^2\Big)\,
\int[\dd\mu_X]\;
\exp\!\Big(\ii\sum_{r=1}^4 k_r\cdot X_N(x_r)\Big).
\label{eq:Io-afterX0}
\end{align}
Only Neumann components $a_i^a$ appear, and the field is linear in them:
\begin{equation}
X_N^a(x)=A_o\sum_{i=1}^N a_i^a f_i(x),
\qquad
A_o^2\sigma_a^2 = 2A^2\sigma_a^2 = \frac{2\alpha'(\Lambda^2-\epsilon^2)}{N}.
\label{eq:Ao2}
\end{equation}
Defining by analogy with the previous section
\begin{equation}
S_i^a \equiv \sum_{r=1}^4 k_r^a f_i(x_r),
\qquad
Q_i \equiv \sum_{r,s=1}^4 (k_r\cdot k_s)\,f_i(x_r)f_i(x_s),
\label{eq:open-SQ}
\end{equation}
so that we have
\begin{equation}
\sum_{r=1}^4 k_r\cdot X_N(x_r)
=
A_o\sum_{i=1}^N\sum_{a=1}^{d_\parallel} a_i^a\,S_i^a,
\qquad
\sum_{a=1}^{d_\parallel}(S_i^a)^2 = Q_i.
\end{equation}
and after the Gaussian average over $\{a_i^a\}$ we are left with
\begin{align}
\Big\langle
\exp\!\Big(\ii\sum_{r=1}^4 k_r\cdot X_N(x_r)\Big)
\Big\rangle_{a}
&=
\exp\!\Big(-\frac{\sigma_a^2A_o^2}{2}\sum_{i=1}^N Q_i\Big)
=
\exp\!\Big(-\frac{\alpha'(\Lambda^2-\epsilon^2)}{N}\sum_{i=1}^N Q_i\Big).
\label{eq:open-aftera}
\end{align}
At this point only the feature average remains.

Introduce a single-neuron random variable
\begin{equation}
Q
\equiv
\sum_{r,s=1}^4 (k_r\cdot k_s)\,f(x_r)f(x_s),
\qquad
f(x)=\frac{\cos(\mathrm{Re}(W)x+c)}{|W|},
\end{equation}
and define
\begin{equation}
\Phi_N \equiv \E_{W,c}\!\left[\exp\!\Big(-\frac{\alpha'(\Lambda^2-\epsilon^2)}{N}Q\Big)\right].
\end{equation}
Since the features are i.i.d.\ across neurons,
\begin{equation}
\int \prod_{i=1}^N\frac{\dd^2W_i}{\Vol(A_{\Lambda,\epsilon})}\frac{\dd c_i}{2\pi}\;
\exp\!\Big(-\frac{\alpha'(\Lambda^2-\epsilon^2)}{N}\sum_{i=1}^N Q_i\Big)
=
\Phi_N^N.
\end{equation}
Expanding at large $N$,
\begin{equation}
\Phi_N
=
1-\frac{\alpha'(\Lambda^2-\epsilon^2)}{N}\,\E_{W,c}[Q]+\cO(N^{-2}),
\end{equation}
we obtain
\begin{equation}
\Phi_N^N \xrightarrow[N\to\infty]{}
\exp\!\Big(-\alpha'(\Lambda^2-\epsilon^2)\,\E_{W,c}[Q]\Big).
\label{eq:open-Phi-limit}
\end{equation}

To evaluate $\E_{W,c}[Q]$, we again average over $c$:
\begin{equation}
\E_c[\cos(u+c)\cos(v+c)] = \tfrac12\cos(u-v),
\qquad
\E_c[\cos^2(u+c)] = \tfrac12.
\end{equation}
This yields
\begin{align}
\E_{W,c}[Q]
&=
\frac12\sum_{r\neq s}(k_r\cdot k_s)\,\E_W\!\left[\frac{\cos(\mathrm{Re}(W)(x_r-x_s))}{|W|^2}\right]
+\frac12\sum_{r=1}^4 k_r^2\,\E_W\!\left[\frac{1}{|W|^2}\right].
\label{eq:EQopen-start}
\end{align}
The $W$-averages reduce to Bessel integrals. With $W=\rho\e^{\ii\theta}$ and $\mathrm{Re}(W)\Delta x=\rho\cos\theta\,\Delta x$,
rotational averaging gives
\begin{align}
\E_W\!\left[\frac{\cos(\mathrm{Re}(W)\Delta x)}{|W|^2}\right]
&=
\frac{2}{\Lambda^2-\epsilon^2}\int_\epsilon^\Lambda\frac{\dd\rho}{\rho}\,J_0(\rho|\Delta x|),
\label{eq:open-bessel}
\\
\E_W\!\left[\frac{1}{|W|^2}\right]
&=
\frac{2}{\Lambda^2-\epsilon^2}\int_\epsilon^\Lambda\frac{\dd\rho}{\rho}
=
\frac{2}{\Lambda^2-\epsilon^2}\log\!\Big(\frac{\Lambda}{\epsilon}\Big).
\label{eq:open-1overW2}
\end{align}
In the scaling window $1/\Lambda\ll |\Delta x|\ll 1/\epsilon$ we use the standard asymptotic
\begin{equation}
\int_\epsilon^\Lambda\frac{\dd\rho}{\rho}\,J_0(\rho|\Delta x|)
=
-\log|\Delta x| + \log\!\Big(\frac{2\Lambda}{\epsilon\e^{\gammaE}}\Big)+\cdots,
\label{eq:open-bessel-asym}
\end{equation}
and hence \eqref{eq:EQopen-start} becomes
\begin{align}
-\alpha'(\Lambda^2-\epsilon^2)\,\E_{W,c}[Q]
&=
2\alpha'\sum_{r<s}(k_r\cdot k_s)\log|x_r-x_s|
-\alpha'\log\!\Big(\frac{2\Lambda}{\epsilon \e^{\gammaE}}\Big)\Big(\sum_{r=1}^4 k_r\Big)^2
\nonumber\\
&\hspace{2.8em}
+\alpha'\log\!\Big(\frac{2}{\e^{\gammaE}}\Big)\sum_{r=1}^4 k_r^2
+\cdots.
\label{eq:open-EQfinal}
\end{align}
Exponentiating and combining with \eqref{eq:Io-afterX0} yields the separation-dependent part of the boundary
correlator,
\begin{equation}
\Big\langle
\exp\!\Big(\ii\sum_{r=1}^4 k_r\cdot X_N(x_r)\Big)
\Big\rangle
=
\prod_{r<s}|x_r-x_s|^{2\alpha'k_r\cdot k_s}
\times \text{(cutoff-dependent constants)}\times (1+\cdots).
\label{eq:open-KN}
\end{equation}
The cutoff-dependent constants in \eqref{eq:open-EQfinal} have two universal pieces: one proportional to
$(\sum_r k_r)^2$, and one proportional to $\sum_r k_r^2$. The $(\sum_r k_r)^2$ term combines naturally with the
Neumann zero-mode factor in \eqref{eq:Io-afterX0}. Defining the effective brane-parallel variance
\begin{equation}
\sigma_{\rm eff,\parallel}^{\prime 2}
\equiv
\sigma_0^2+2\alpha'\log\!\Big(\frac{2\Lambda}{\epsilon \e^{\gammaE}}\Big),
\label{eq:sigmaeff-open}
\end{equation}
we can rewrite the product as
\begin{equation}
\exp\!\Big(-\frac{\sigma_0^2}{2}\Big(\sum_r k_r\Big)^2\Big)\;
\exp\!\Big(-\alpha'\log\!\Big(\frac{2\Lambda}{\epsilon \e^{\gammaE}}\Big)\Big(\sum_r k_r\Big)^2\Big)
=
\exp\!\Big(-\frac{\sigma_{\rm eff,\parallel}^{\prime 2}}{2}\Big(\sum_r k_r\Big)^2\Big).
\end{equation}
The remaining factor proportional to $\sum_r k_r^2$ is removed by the multiplicative renormalization
$Z_o(k)$, which implements the usual normal-ordering logic for boundary exponentials and supplies the
Gaussian normalization required for the momentum-conservation limit. A convenient choice is
\begin{equation}
Z_o(k)
=
\Big(\frac{\e^{\gammaE}}{2}\Big)^{\alpha'k^2}
\Big(\frac{\sigma_{\rm eff,\parallel}^{\prime 2}}{2\pi}\Big)^{\frac{d_\parallel}{8}}
=
\Big(\frac{\e^{\gammaE}}{2}\Big)^{\alpha'k^2}
\Big(\frac{\sigma_{\rm eff,\parallel}^{\prime 2}}{2\pi}\Big)^{\frac{p+1}{8}}.
\label{eq:Zo}
\end{equation}
Collecting the pieces, the renormalized four-tachyon disk amplitude on the neural D$p$-brane becomes
\begin{align}
&A^{(4)}_{Dp}(p_1,p_2,p_3,p_4;Y)
=
\frac{T_p^{\rm NN}\,g_s}{\Vol(SL(2,\mathbb R))}\;
\e^{\ii Y\cdot\sum_{r=1}^4 q_r}\;
\Big(\frac{\sigma_{\rm eff,\parallel}^{\prime 2}}{2\pi}\Big)^{\frac{p+1}{2}}\;\nonumber\\
&\times\exp\!\Big(-\frac{\sigma_{\rm eff,\parallel}^{\prime 2}}{2}\Big(\sum_{r=1}^4 k_r\Big)^2\Big)
\int \prod_{r=1}^4 \dd x_r\;
\prod_{r<s}|x_r-x_s|^{2\alpha'k_r\cdot k_s},
\label{eq:open-final}
\end{align}
up to subleading corrections outside the scaling regime \eqref{eq:scaling-window} and $1/N$ effects.
In the limit $\sigma_{\rm eff,\parallel}^\prime\to\infty$, the Gaussian becomes a delta function enforcing
momentum conservation along the brane,
\begin{equation}
\Big(\frac{\sigma_{\rm eff,\parallel}^{\prime 2}}{2\pi}\Big)^{\frac{p+1}{2}}
\exp\!\Big(-\frac{\sigma_{\rm eff,\parallel}^{\prime 2}}{2}\Big(\sum_r k_r\Big)^2\Big)
\xrightarrow[\sigma_{\rm eff,\parallel}^\prime\to\infty]{}
\delta^{(p+1)}\!\Big(\sum_{r=1}^4 k_r\Big),
\end{equation}
while no transverse delta function arises because the Dirichlet zero modes are frozen. For physical open
string states on a single D$p$-brane one typically has $q_r=0$, in which case the phase
$\e^{\ii Y\cdot\sum q_r}$ is absent.

Finally, we stress again that the only genuinely new datum beyond the boundary logarithmic kernel is the
overall constant $T_p^{\rm NN}$. Once it is fixed by a single normalization choice, it multiplies all disk
amplitudes with D$p$ boundary conditions, exactly as a D$p$-brane tension does in the conventional worldsheet
formalism.

\section{Conclusions and outlook}

In this work we extended the neural-network rewriting of the genus-zero worldsheet path integral initiated in \cite{Frank:2026nnstring} to two settings that probe genuinely new structure beyond tachyon correlators on closed worldsheets. The first is the inclusion of excited closed-string insertions, where the basic issue is not the exponential sector---whose cutoff-dependent self-contractions are removed by the usual multiplicative renormalization---but rather the definition of derivative composites such as $\partial X^\mu\bar\partial X^\nu$. In a neural regularization this composite carries an additional ultraviolet-sensitive local term, the parameter-space counterpart of the coincident contraction in the continuum CFT. Our central technical input was a renormalization prescription tailored to the ensemble computation: Gaussian normal ordering with respect to the output-weight measure at fixed random features. This subtraction is exact at finite width and fixed features, and it cleanly removes the contact term while leaving intact the physically meaningful contractions against exponentials. Implementing this prescription, we derived the sphere four-point matter integrand with one level-$(1,1)$ insertion and three tachyons and recovered the expected rational prefactor multiplying the Koba--Nielsen factor, including the correct $\alpha'$-dependence fixed by the neural scaling.

 The second extension concerned worldsheets with boundary and mixed Neumann/Dirichlet conditions. We proposed a neural realization of D$p$ boundary conditions via an image-like construction acting directly on the bulk neural field. On the boundary this yields fluctuations only along Neumann directions and frozen Dirichlet components fixed to a brane position $Y^i$, reproducing the standard kinematic structure of open-string vertices. Evaluating the resulting disk correlator at large width, we recovered the boundary logarithmic kernel and hence the boundary Koba--Nielsen factor, together with the emergence of momentum conservation along the brane from an explicit Neumann zero-mode integral in the appropriate infinite-variance limit.

There are several natural directions for further development. On the technical side, it would be valuable to systematize the treatment of more general composite operators and higher excited states, as well as to incorporate superstring sectors where fermions and picture-changing introduce additional layers of regularization and renormalization. It would also be interesting to compute mixed open--closed amplitudes on the disk in the same ensemble, clarifying the neural counterpart of boundary states and the standard closed-string couplings to D-branes. More broadly, understanding finite-width corrections---which in the NN-FT viewpoint correspond to controlled deviations from the strict free-field limit---may provide a new handle on worldsheet interactions and, potentially, on contributions beyond leading genus. We hope that the explicit parameter-space representation emphasized here will make such questions concrete, both analytically and numerically, and will continue to sharpen the emerging dictionary between neural ensembles and worldsheet field theory suggested by \cite{Frank:2026nnstring}.

\end{document}